\documentclass[conference]{IEEEtran}
\IEEEoverridecommandlockouts
\usepackage{cite}
\usepackage{amsmath,amssymb,amsfonts}
\usepackage{algorithmic}
\usepackage{graphicx}
\usepackage{textcomp}
\usepackage{xcolor}

\def\BibTeX{{\rm B\kern-.05em{\sc i\kern-.025em b}\kern-.08em
    T\kern-.1667em\lower.7ex\hbox{E}\kern-.125emX}}
\begin{document}

\title{Scope of Online Maternal Health Information in Kinyarwanda and Opportunities for Digital Health Developers
}

\author{\IEEEauthorblockN{1\textsuperscript{st} Joselyne Muragijemariya}
\IEEEauthorblockA{\textit{Carnegie Mellon University Africa} \\
Kigali, Rwanda \\
jmuragij@alumni.cmu.edu}
\and
\IEEEauthorblockN{2\textsuperscript{nd} Valentine Ihogoza}
\IEEEauthorblockA{\textit{Carnegie Mellon University Africa} \\
Kigali, Rwanda \\
vihogoza@alumni.cmu.edu}
\and
\IEEEauthorblockN{3\textsuperscript{rd} Edith Talina Luhanga}
\IEEEauthorblockA{\textit{Carnegie Mellon University Africa} \\
Kigali, Rwanda \\
eluhanga@andrew.cmu.edu}
}

\maketitle

\begin{abstract}
Maternal health literacy is associated with greater odds of positive pregnancy outcomes. There is an increasing proliferation of websites dedicated to maternal health education, but the scope and quality of their content varies widely. In this study, we analyzed the main topics covered on maternal health websites that offer content in the low-resource Kinyarwanda language (mainly spoken by 12 million Rwandans). We used web scraping to identify maternal health websites. We utilized a topic modeling, using the Non-Negative Matrix Factorization (NMF) algorithm, to identify the topics. We found five main topics: (1) pregnancy danger signs, (2) child care, (3) intimacy (sex), (4) nutrition, and (5) the importance of doctor consultations. However, the articles were short and did not cater to fathers, pregnant adolescents, or those experiencing gender-based violence (GBV) or mental health challenges. This is despite 12.5\% women of reproductive age in Rwanda being victims of GBV and one in five women in low- and middle-income countries experiencing mental illness during the perinatal period. We recommend three automated tools, a topic recommender tool, culturally relevant automated articles, and website quality check tools, to guide software and health content developers.\end{abstract}


\section{Introduction}
Africa continues to have the world’s highest rates of maternal deaths (70\% of global deaths) \cite{whoglobal}. In Rwanda, there were 203 deaths per 100,000 live births in 2019/2020 \cite{dhsrwanda2020}, with the leading causes of death being sepsis, hemorrhage, and hypertensive disorders \cite{rulisa2021causes}. These three are also among the leading causes of maternal deaths worldwide \cite{whoglobal}.

In many countries in Africa, maternal deaths occur because women seek care too late to receive lifesaving treatment (phase 1 delays) \cite{mgawadere2017factors}\cite{ nassoro2020maternal}\cite{mohammed2020ten}, and this is usually due to a lack of knowledge of obstetric danger signs \cite{roberts2021current}\cite{alobo2021research}. Unsurprisingly, a panel of more than 900 experts identified the improvement of maternal knowledge about danger signs as one of the most priority interventions to address maternal mortality in the Sub-Saharan Africa (SSA) region \cite{alobo2021research}. 

Maternal counseling services provided as part of antenatal care (ANC) are one way to promote maternal knowledge of danger signs. The World Health Organization (WHO) recommends a minimum of eight ANC contacts (one in the first trimester, two in the second trimester, and five in the third trimester) \cite{tunccalp2017recommendations}, but in SSA, only half of pregnant women (55\% in Eastern and Southern Africa and 56\% in Western and Central Africa) manage four visits to ANC during pregnancy \cite{unicefmaternaldeaths}. Poor adoption of ANC is due to long distances to ANC facilities \cite{dusingizimana2023predictors}, difficulty in getting money for treatment \cite{ahinkorah2021examining}, spousal reluctance to accompany coupled women despite clinics implementing a spousal accompaniment policy \cite{mgata2019factors, maluka2020pregnant, ahinkorah2021examining}, lack of knowledge about the importance of ANC \cite{maluka2020pregnant} or when to start ANC \cite{mgata2019factors}, and sometimes poor quality of ANC services \cite{maluka2020pregnant}. There is also a low demand for ANC services among some groups, e.g. women of lower education levels \cite{smith2019demand}, women with unwanted pregnancies \cite{smith2019demand, bobo2021poor}, women with high parity \cite{smith2019demand, sserwanja2022factors}, and working women \cite{sserwanja2022factors}.  

Various countries have launched patient-facing digital applications for maternal health to counteract the ANC utilization gap and its effect on maternal knowledge. These include MomConnect in South Africa \cite{barron2018mobile} and GiftedMom in Cameroon and Nigeria \cite{temgoua2018innovative}. The applications provide short message service (SMS) or web- and app-based pregnancy information, reminders to attend ANC clinics, and even collect feedback on ANC quality. 

Urban women also are increasingly using health and pregnancy websites and apps (informal mHealth), available primarily in English, for their maternal health information needs \cite{obasola2017women, kaaya2021maternal, khamis2023maternal}. Websites generally have more information than other media sources and can be accessed anywhere, anytime, without installation \cite{obasola2017women}. However, accessing health information in a second language can be frustrating and difficult for users \cite{kaaya2021maternal, chu2022impact}. Furthermore, users who prefer to access health information in their native language are less likely to use second language information resources, even if they are proficient in the second language \cite{massey2017nativity, khoong2019health}. Thus, to promote online health information search, it is essential to ensure that sufficient high-quality information resources are available in local languages.  

This study aimed to evaluate the amount and type of maternal health information available online in the Kinyarwanda language, which is spoken by more than 99\% of the Rwandan population \cite{languagesrwanda2}. We used Web screaming tools to extract data from pregnancy web pages. We then applied topic modeling to identify the unique topics covered in the retrieved Web pages. We found five main topics: (1) Preventive measures for safe pregnancy, (2) Child care during illness, (3) intimacy (sex) during the antenatal period, (4) healthy diet during the antenatal period, and (5) importance of doctor consultation. However, the articles were fairly short (less than 3000 words) and therefore did not fully cover the main aspects of the presented topic. We identify five important underexplored topics and recommend the development of automated topic recommendation and quality assessment tools to guide software and health content developers to produce more useful and usable websites. 

\section{Related work}
\subsection{Content and Quality of Online Health Information Resources}
Numerous studies have evaluated the content on pregnancy and birth Web pages, websites, and apps. An evaluation of 126 English and Spanish pregnancy websites against the LIDA tool found only two websites with excellent reliability. Almost half (49.2\%) had poor or very poor scores \cite{artieta2018evaluation}. Virani et al.\cite{virani2019parenting} used the Mobile Application Rating Scale (MARS) to evaluate free pregnancy apps recovered from Google Play. The main categories of applications found were (1) tracking apps, which allow parents to record the times babies eat, sleep, or have bowel movements; (2) informational apps (3) photo sharing apps; and (4) sleeping aid apps, which provided features such as lullabies. Medication applications were also found. The applications were generally medium to high quality, with MARS scores for the 16 apps reviewed ranging from 4.2 to 4.8. However, topic-specific reviews show low to average information quality. Pregnancy nutrition is one of the most studied topics. Storr et al. \cite{storr2017online} reviewed 693 websites and compared the content with the 2013 Australian Dietary Guidelines. Only 39.5\% of websites contained accurate information, while the rest contained mixed or inaccurate information. Another study on 18 Australian government and industry websites \cite{cannon2020review} found that none adhered to the ``Healthy Eating during Pregnancy'' guidelines produced by the Australian government. Not adequately covered topics included healthy weight gain during pregnancy, limiting sugar and saturated fat intake, and drinking water during pregnancy. Other topics such as obesity in pregnancy \cite{al2016online}, physical activity during pregnancy \cite{hayman2021quality}, contraception \cite{swartzendruber2018contraceptive}, and malaria in pregnancy \cite{hamwela2018evaluation} similarly have information ranging from poor to medium quality. 

Surveys with women who use online pregnancy information reinforce the lack of high-quality sources. In \cite{vogels2022sources}, Dutch women rated online health information resources as less trustworthy than information from peers and health professionals, while in \cite{ojaperv2023pregnancy}, 77/7\% of Estonian women reported that online health information was sometimes confusing and controversial. Despite this, online sources were the main source of pregnancy information among Estonian women, and information was mainly sought for decision making \cite{ojaperv2023pregnancy}. Analysis of posts on pregnancy and birth forums also reveals unmet information needs. On the \textbf{WhatToExpect} forum, for example, the discussions mainly revolved around concerns of miscarriage in the first trimester and labor in the third trimester. Other discussions included concerns about ultrasound results and fetal movements and questions about how a new baby might change family dynamics \cite{wexler2020pregnancy}. Pregnant adolescents, on the other hand, primarily searched for information on forums. Their questions included whether a missed period might be due to pregnancy or participation in sports during pregnancy \cite{bostwick2019could}. Information obtained from peers in such forums can be risky. In \cite{wigginton2017safe}, questions about the use of e-cigarettes in pregnancy sometimes received responses suggesting that e-cigarettes are a safer and more acceptable alternative to quitting cold turkey or using normal cigarettes. 

\subsection{Topic Modeling and Its Applications}
Up to 7\% of Google queries (70,000 every minute) are related to health \cite{drGoogle}. Such data can reveal both personal and population-level health status. In \cite{asch2019google}, the authors found that health information searches of patients who visited an emergency department doubled in the week before the visit. Topic modeling is an unsupervised machine learning technique that summarizes the content of documents into distinct groups of co-occurring words (topics) \cite{boyd2017applications}. In health and biosciences, topic modeling has been used to identify topics in discussion forums \cite{zou2018analyzing, smoll2021barriers, qin2022exploring}, research trends \cite{zou2018analyzing}, classifying electronic health records according to the International Classification of Diseases \cite{lebena2022preliminary}, and personalized disease management\cite{ni2022topic}. The modeler pre-specifies the number of topics \textbf{K}, and the optimal value can be assessed by evaluating the perplexity, i.e., how well the model predicts unseen data and the model fitting time \cite{ni2022topic}. 

\section{Methods}
\subsection{Keyword selection}
We selected ''pregnancy'', ''conception'', and ''maternal health'' as the primary keywords and translated them into Kinyarwanda. We used the ''Answer the Public'' tool to confirm which keywords were widely used by looking at the search volume of each term. We found ''gutwita'' (pregnancy), ''gusama'' (conceiving), ''atwite'' (pregnant), ''utwite'' (pregnant), and ''batwite'' (pregnant women) to be the main terms used online. These were selected as the final keywords.

\subsection{Data Collection}
People often use search engines to search for health information rather than going directly to health websites \cite{lagan2011impact,kaaya2021maternal,lee2022topics}. We therefore used a web scraper to automate the retrieval and extraction of data from websites returned by search engines. The scraping was carried out in March 2024. 

Each of the selected keywords was entered into the scraper. The scraper sent this search term to www.google.com, since this is the most widely used search engine in Africa \cite{googlemarket}, and returned results from the first three pages. We limit the number of result pages to three, following online quality information assessment tools such as Health Information Website Evaluation Tool (HIWET) \cite{zubiena2022development}. The scraper returned website URLs, the number of web pages, the URLs of each web page, and the contents of each web page. The data were extracted in a CSV file. The total number of web pages in this initial dataset was 3,402. 

\subsection{Website Screening}
We analyzed the titles and main paragraphs of the scraped websites and included them in the final dataset if they were (1) written in Kinyarwanda and (2) had content related to pregnancy. We excluded websites that (1) were inaccessible due to a paywall or required log-in, (2) contained only videos, since these are excluded in HIWET \cite{zubiena2023development}, and (3) were duplicates of websites already included. After screening, the final websites were exported as a CSV file containing the URL and scraped data. The final dataset contained 2,642 unique web pages. 

\subsection{Data Cleaning and Preprocessing}
The dataset was cleaned by removing punctuation, English and French words, numerical values, Kinyarwanda stop words, and short words (3 characters or less). We then applied stemming techniques to the dataset. 

After cleaning the data, the text content of the pages was tokenized using the NLTK library, breaking down each document into individual words (tokens). Subsequently, the tokenized text was converted into a bag-of-words (BoW) representation using the CountVectorizer from the scikit-learn package in Python. This process involves transforming the text data into a matrix where each row corresponds to a document, and each column represents a unique word in the entire corpus. We tabulated the frequencies of words \ref{table:common_words} to identify emerging themes in the documents. 

\begin{table}[h!]
\centering
\begin{tabular}{|c|c|c|}
\hline
\textbf{Kinyarwanda} & \textbf{English} & \textbf{Frequency} \\
\hline
umwana & child & 384 \\
\hline
umugore & woman & 367 \\
\hline
utwite & pregnant & 323 \\
\hline
inda & pregnancy & 232 \\
\hline
igihe & time & 168 \\
\hline
neza & well & 141 \\
\hline
atwite & pregnant & 138 \\
\hline
kugira & having & 138 \\
\hline
gutwita & pregnancy & 133 \\
\hline
umubyeyi & parent & 130 \\
\hline
muganga & doctor & 126 \\
\hline
mbere & before & 121 \\
\hline
batwite & pregnant women & 109 \\
\hline
nyuma & after & 101 \\
\hline
imibonano & intimacy & 100 \\
\hline
abagore & women & 96 \\
\hline
gukora & doing & 88 \\
\hline
mpuzabitsina & sexual & 88 \\
\hline
ubuzima & health & 86 \\
\hline
kubyara & giving birth & 86 \\
\hline
\end{tabular}
\caption{Most Common Words in Kinyarwanda maternal health web pages with their English translations and frequencies of occurrence}
\label{table:common_words}
\end{table}

\subsection{Model Training and Evaluation}
We used Non-Negative Matrix Factorization (NMF) and Latent Dirichlet Allocation (LDA) to discover the topics in the dataset. Latent Dirichlet Allocation (LDA) is a probabilistic model that represents topics as distributions over words, using probabilities to assess the likelihood that each word belongs to a specific topic. This framework enables documents to be characterized as mixtures of multiple topics, revealing the complex thematic structure often found in written texts.
\cite{blei2003latent}. In contrast, NMF uses linear algebra to uncover topics \cite{lee1999learning}. Although LDA has traditionally received more attention, some studies have found that NMF provides more coherence and generality \cite{o2015analysis, latif2021analyzing}. We therefore investigated its performance against LDA for the Kinyarwanda dataset. The Gemsim library was used to implement the models. 

We used the Coherence Score to evaluate the interpretability of the results and cosine similarity to evaluate diversity. Topic coherence measures how meaningful a topic is by assessing the frequency with which the top words identified in a topic appear together in documents and the distinctiveness of the topics identified by the individual models. The Coherence Score ranges from 0 to 1 (1 being the highest coherence), and a higher score indicates that the words in a topic frequently co-occur in the documents and are therefore more interpretable \cite{kherwa2019topic}. Topic diversity assesses how distinct the identified topics are from each other. We used a cosine similarity matrix to identify pairwise similarity between different topics. Cosine similarity scores range from -1 to 1, with a 0 score indicating no similarity and a score of 1 indicating perfect similarity.

\begin{figure}
    \centering
    \includegraphics[width=0.7\linewidth]{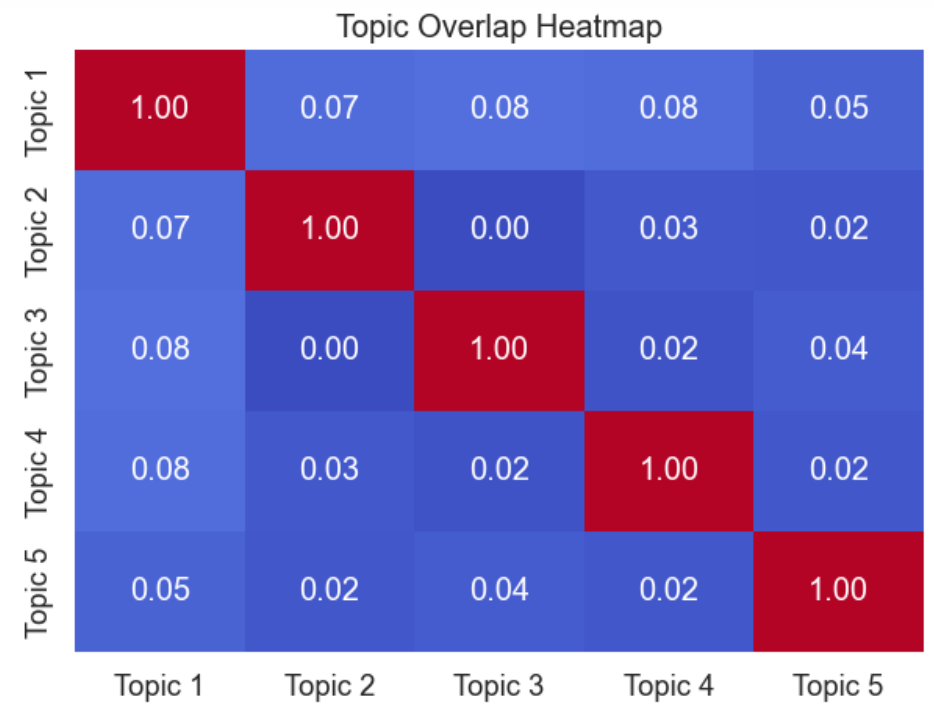}
    \caption{Cosine Similarity Matrix for the 5 NMF  Topics}
    \label{fig:cos-sim}
\end{figure}

Both LDA and NMF require that the number of topics be specified in advance. We selected 4, 5, 7, and 10 topics for training. The models with five topics had the highest coherence and Hellinger distance scores. A manual analysis of the topics and content by the first authors verified that five topics yielded more coherent results for both models, and the NMF had higher coherence scores than LDA across the various topics (Table \ref{table:coherence}). The results section, therefore, focuses on explaining the five topics revealed with topic modeling using the NMF model. The cosine similarity matrix for the five topics identified by the NMF model is shown in Fig \ref{fig:cos-sim}. 

\begin{table}[h!]
\centering
\begin{tabular}{|c|c|c|}
\hline
\textbf{No. of Topics} & \textbf{NMF} & \textbf{LDA} \\ \hline
4 & 0.62 & 0.51 \\ \hline
5 & 0.71 & 0.54 \\ \hline
7 & 0.70 & 0.50 \\ \hline
10 & 0.69 & 0.48 \\ \hline
 \end{tabular}
\caption{Coherence scores for the NMF and LDA models on 4, 5, 7, and 10 topics.}
\label{table:coherence}
\end{table}

\section{Results}
The five topics identified by the NMF model were on ''preventive measures for safe pregnancy'', ''child care'', ''intimacy (sex)'', ''healthy diet '', and ''importance of doctor consultation''. The coherence score achieved was 0.70. Table \ref{table:NMF_topics} outlines the Kinyarwanda words in each topic and their translations. 

\begin{table*}[t]
\caption{The five topics (and corresponding words) identified by the NMF model}
\begin{tabular}{p{0.2\linewidth}| p{0.3\linewidth} | p{0.3\linewidth}}
\hline
\textbf{Topic} & \textbf{Words (Kinyarwanda) } & \textbf{Words (English)} \\
\hline
\textbf{Topic 1:} Preventive measures for safe pregnancy & umugore, utwite, inda, yilinda, umwana, ufite, kubyara, akareba, kureba, ikintu & woman, pregnant, abodmen, protected, child, having, giving birth, watching, watching something \\
\hline
\textbf{Topic 2:} Child care during illness & umwana, umuriro, kubuzima, umubyeyi, ashobora, agatabo, imiti, kenshi, ibisobanuro, mwana & child, fever, for health, parent, can, booklet, medication, often, explanation, child \\
\hline
\textbf{Topic 3:} Intimacy (sex) during the antenatal period & imibonano, mpuzabitsina, gukora, uburyo, umugore, nyuma, idakingiye, umugabo, atwite, bikaba & intercourse, sex, work, method, woman, after, unprotected, man, pregnant, being \\
\hline
\textbf{Topic 4:} Healthy diet during the antenatal period & umubyeyi, utwite, kugira, neza, agomba, atwite, kurya, ubuzima, munsi, ingenzi & mother, pregnant, have, well, should, pregnant, eat, health, under, important \\
\hline
\textbf{Topic 5:} Doctor consultation & igihe, gutwita, mbere, inda, muganga, abagore, batwite, urubyaro, buryo, abana & time, pregnancy, before, pregnancy, doctor, women, pregnant, offspring, method, children \\ 
\hline
\end{tabular}
\label{table:NMF_topics}
\end{table*}

\begin{figure}
    \centering
    \includegraphics[width=0.7\linewidth]{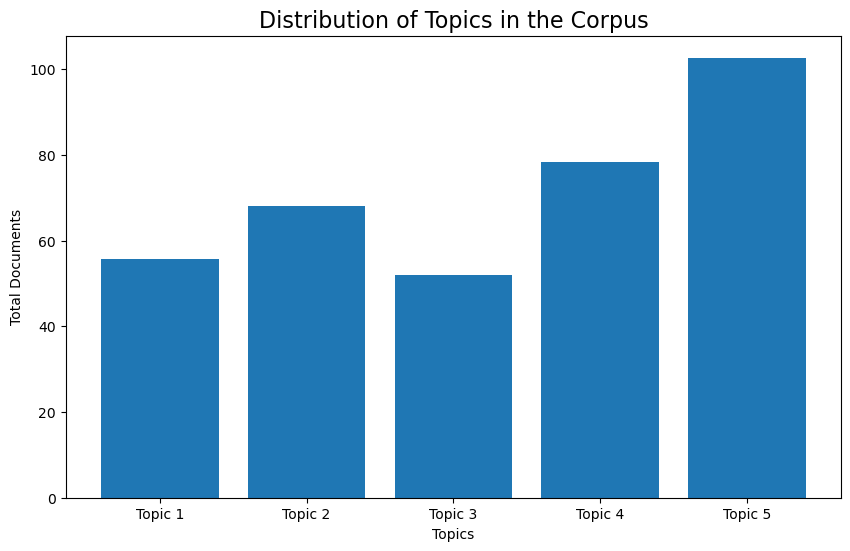}
    \caption{Caption}
    \label{fig:topicdist}
\end{figure}

The proportion of documents that talked about each topic is shown in \ref{fig:topicdist}. The importance of doctor consultations and healthy diets were the two most discussed topics, while intimacy during pregnancy was the least discussed. 

\textbf{Topic 1: Preventive Measures for Safe Pregnancy}
Topic 1's key terms were ''umugore'' (woman), ''utwite'' (pregnant), ''inda'' (pregnancy), ''yilinda'' (protecting), ''umwana'' (child), ''ufite'' (having), ''kubyara'' (giving birth), ''akareba'' (watching), ''kureba'' (looking), and ''ikintu'' (thing). Analysis of the documents that contain the topic revealed that they emphasized the importance of educating expectant mothers on protecting their pregnancies and ensuring safe delivery. The primary measures highlighted were guidelines for regular prenatal check-ups, avoiding harmful substances such as alcohol, and recognizing the danger signs of complications. The main point was to empower pregnant women with the knowledge and resources necessary to protect their health and their unborn children.\\

\textbf{Topic 2: Child care during illness}
The topic had terms like ''umwana'' (child), ''umuriro'' (fever), ''kubuzima'' (health), ''umubyeyi'' (parent), ''ashobora'' (can), ''agatabo'' (booklet), ''imiti'' (medicine), ''kenshi'' (often), ''ibisobanuro'' (explanations), and ''mwana'' (child). The documents provided guidance on how to address children's health issues, including managing common illnesses, ensuring proper nutrition, and administering medications safely. \\

\textbf{Topic 3: Intimacy (sex) during the Antenatal Period}
The keywords comprised in this topic included ''imibonano'' (intimacy), ''mpuzabitsina'' (sexual), ''gukora'' (doing), ''uburyo'' (ways), ''umugore'' (woman), ''nyuma'' (after), ''idakingiye'' (unprotected), ''umugabo'' (man), ''atwite'' (pregnant), and ''bikaba'' (becoming). The documents covering this topic underscored the importance of open communication between partners and provided guidance on maintaining a healthy sexual relationship during the antenatal period. The topic also covered general topics of sexual health education, such as the importance of protection and the implications of unprotected sex. \\

\textbf{Topic 4: Healthy Diet during the Antenatal Period}
This topic used terms such as "umubyeyi" (parent), "utwite" (pregnant), "kugira" (having), "neza" (well), "agomba" (must), "kurya" (eating), "ubuzima" (health), "munsi" (under), and "ingenzi" (important), to emphasize the importance of eating a healthy diet during pregnancy. This was done by highlighting the critical role of nutrition in the health of both the mother and the developing fetus and by providing information on nutritional and wellness needs of pregnant women. Guidance on how to provide a healthy diet was also provided. \\

\textbf{Topic 5: Importance of Consultation with Doctors}
Finally, the last topic focused on the importance of consulting with healthcare professionals during pregnancy, with terms like "igihe" (time), "gutwita" (pregnancy), "muganga" (doctor), "abagore" (women), "batwite" (pregnant), "urubyaro" (progeny), "bwo" (ways), and "abana" (children). This topic centers around prenatal and postnatal care, highlighting the importance of medical check-ups before and after childbirth. It also includes information about the different stages of pregnancy and the necessary preparations to welcome a new child.\\\\

\subsection{Discussion}
Our study accessed 2,642 unique web pages on maternal health written in Kinyarwanda. The topics covered on Kinyarwanda maternal health webpages focus on both the mother's health and well-being during the antenatal period and the child's health after birth. Maternal health topics focused on education on obstetric danger signs and preventing complications (Topic 1), safe sex (Topic 3), healthy diet (Topic 4), and the importance of consulting health workers (Topic 5). The child health topic (Topic 2) focused on healthy diets and managing illnesses. 

The evaluation of maternal health information published by Rwandan print media between January 2008 and March 2018 found 342 articles, mainly on events and conferences \cite{gugsa2016newspaper}. The higher volume of articles on the Internet and their focus on maternal education indicate the increased use of media technologies for maternal health in Rwanda over the past decade. Globally, the most searched-for pregnancy topics are fetal development, pregnancy symptoms and complications, and nutrition \cite{conrad2024health}. These topics were among the top five topics found online in Kinyarwanda. Nurses and midwives in Rwanda also emphasize maternal nutrition and pregnancy complications (what to do if a serious problem occurs) \cite{rurangirwa2018quality} as well as family planning during ANC sessions. However, other topics that they cover, such as prevention of sexually transmitted diseases and place of delivery, are not sufficiently covered online. 

\subsection{Unaddressed Information Gaps}
We found several other information gaps. These include limited information on (1) preconception and postpartum care, (2) fathers' roles in pregnancy and childbirth, (3) gender-based violence during pregnancy, (4) unique needs of pregnant adolescents, and (5) perinatal mental health. These gaps were also identified in \cite{artieta2018evaluation}\cite{swartzendruber2018contraceptive}\cite{mniszak2020nothing}\cite{onyeze2020experiences} and are vital to address for the following reasons: 
\begin{itemize}
\item The imbalance between the information provided to expecting mothers versus fathers can perpetuate social norms around gender roles. These, in turn, leave women to bear the burden of pregnancy and childcare alone \cite{mniszak2020nothing}\cite{onyeze2020experiences}. 
\item In Rwanda, 12. 5\% of women of reproductive age in Rwanda have experienced gender-based violence (GBV) \cite{nuwabaine2023sexual}. Women who have experienced intimate partner violence (IPV) in Rwanda are less likely to initiate timely ANC care and are less likely to have four ANC visits \cite{bahati2021intimate}. GBV / IPV during pregnancy is also associated with an increased risk of maternal mental health problems such as depression and anxiety \cite{bahati2021intimate}. The limited focus on this topic online is a significant gap as maternal community health workers in Rwanda have also expressed insufficient knowledge on how to handle disclosures of violence and abuse \cite{nkurunziza2024integrating} and only \cite{rurangirwa2018quality}  7. 8\% of midwives and nurses in Rwanda discuss GBV during ANC sessions. 
\item One in five women in low- and middle-income countries experiences mental ill health, usually depression or anxiety, during the perinatal period \cite{world2022guide} and poor mental health during pregnancy can lead to reduced use of ANC and is associated with complications such as preeclampsia, hemorrhage, and premature delivery \cite{world2022guide}.
\end{itemize}

\subsection{Recommended Digital Tools for Improved Maternal Health Website and Web Content Design}
We call for the development of technologies to guide software developers and creators of health content in developing more useful and usable websites. The tools can include the following. 
\begin{itemize}
    \item \textbf{Topic recommender systems: }There is a wealth of information available on maternal and paternal information needs, the scope of maternal health content on applications and websites, and on priority maternal health interventions for various countries according to country strategic plans or demographic and health survey data. Search engines like Google also publish data on top health-related queries received in various geographical regions \cite{https://about.google/stories/year-in-search/}. Natural language processing techniques can be used to extract and summarize important topics by geographical region, year by year, and highlight information gaps that content creators should address. 
    \item \textbf{Automated health quality scoring tools: } Various quality assessment questionnaires have been developed for health information. These include the LIDA tool, DISCERN questionnaire on treatment options \cite{charnock1999discern} and the HIWET tool to evaluate website information \cite{zubiena2023development}. Plugins for content management systems (CMS) could be developed to automatically score content based on these questionnaires and to provide actionable feedback on how to improve the content. 
    \item \textbf{Automaed usability and credibility assessment tools: } several existing tools can automatically assess the accessibility and usability of websites and apps. These include Siteimprove, WAVE, UsabilityHub, and Google Lighthouse. However, the theoretical usability frameworks used by these tools vary widely and are not always clear \cite{namoun2021review}. Usability also depends on sociocultural factors \cite{alexander2021influence}, and these existing tools do not always capture them. There is increasing interest in the use of artificial intelligence, including the use of digital humans and synthetic users, for user experience (UX) design \cite{stige2024artificial}. We call for data-driven development of synthetic personas from varied socio-economic, cultural, and geographical backgrounds based on popular usability assessment metrics, like the System Usability Scale, Mobile Application Rating Scale, and Norman's ten heuristics. Automated usability tools could then leverage these personas to flag structural and content gaps or errors on websites and apps, with actionable feedback on improving site performance according to the selected metric.      
\end{itemize}

\subsection{Study Limitations}
This study only analyzed the web pages in the first three Google search results and provided textual information rather than image or video-based information. However, previous studies show a high preference for social media and ''TikTok'' style videos \cite{lee2022topics} among pregnant women. The study also analyzed topics without assessing the readability, timeliness (when last updated) or general quality of the information. 

\section{Conclusion}
This study outlined the five main maternal health topics discussed in Kinyarwanda health websites. The topics cover the antenatal period, focusing on vigilance in monitoring obstetric danger signs, child care, intimacy, good nutrition habits, and doctor consultations. We outline unmet information needs, such as gender-based roles, violence, and mental illness during pregnancy. We call for additional research to assess the quality of health information and recommend three types of digital applications to guide software and health content developers to build more useful and usable websites. 


\begin{thebibliography}{10}
\providecommand{\url}[1]{#1}
\csname url@samestyle\endcsname
\providecommand{\newblock}{\relax}
\providecommand{\bibinfo}[2]{#2}
\providecommand{\BIBentrySTDinterwordspacing}{\spaceskip=0pt\relax}
\providecommand{\BIBentryALTinterwordstretchfactor}{4}
\providecommand{\BIBentryALTinterwordspacing}{\spaceskip=\fontdimen2\font plus
\BIBentryALTinterwordstretchfactor\fontdimen3\font minus \fontdimen4\font\relax}
\providecommand{\BIBforeignlanguage}[2]{{%
\expandafter\ifx\csname l@#1\endcsname\relax
\typeout{** WARNING: IEEEtran.bst: No hyphenation pattern has been}%
\typeout{** loaded for the language `#1'. Using the pattern for}%
\typeout{** the default language instead.}%
\else
\language=\csname l@#1\endcsname
\fi
#2}}
\providecommand{\BIBdecl}{\relax}
\BIBdecl

\bibitem{whoglobal}
\BIBentryALTinterwordspacing
W.~H. Organization, ``Maternal mortality,'' 2024, accessed: 26 April 2024. [Online]. Available: \url{https://www.who.int/news-room/fact-sheets/detail/maternal-mortality}
\BIBentrySTDinterwordspacing

\bibitem{dhsrwanda2020}
N.~I. of~Statistics~of Rwanda~(NISR), M.~of~Health~(MOH), and ICF, ``Rwanda demographic and health survey 2019-20 final report,'' Tech. Rep., 2020.

\bibitem{rulisa2021causes}
S.~Rulisa, P.~Ntihinyurwa, D.~Ntirushwa, A.~Wong, and A.~Olufolabi, ``Causes of maternal mortality in rwanda, 2017--2019,'' \emph{Obstetrics \& Gynecology}, vol. 138, no.~4, pp. 552--556, 2021.

\bibitem{mgawadere2017factors}
F.~Mgawadere, R.~Unkels, A.~Kazembe, and N.~van~den Broek, ``Factors associated with maternal mortality in malawi: application of the three delays model,'' \emph{BMC pregnancy and childbirth}, vol.~17, pp. 1--9, 2017.

\bibitem{nassoro2020maternal}
M.~M. Nassoro, P.~Chetto, E.~Chiwanga, A.~Lilungulu, D.~Bintabara, J.~Wambura \emph{et~al.}, ``Maternal mortality in dodoma regional referral hospital, tanzania,'' \emph{International Journal of Reproductive Medicine}, vol. 2020, 2020.

\bibitem{mohammed2020ten}
M.~M. Mohammed, S.~El~Gelany, A.~R. Eladwy, E.~I. Ali, M.~T. Gadelrab, E.~M. Ibrahim, E.~M. Khalifa, A.~K. Abdelhakium, H.~Fares, A.~M. Yousef \emph{et~al.}, ``A ten year analysis of maternal deaths in a tertiary hospital using the three delays model,'' \emph{BMC Pregnancy and Childbirth}, vol.~20, pp. 1--8, 2020.

\bibitem{roberts2021current}
E.~Roberts, S.~M. Baird, and S.~Martin, ``Current key challenges in managing maternal sepsis,'' \emph{The Journal of Perinatal \& Neonatal Nursing}, vol.~35, no.~2, pp. 132--141, 2021.

\bibitem{alobo2021research}
M.~Alobo, C.~Mgone, J.~Lawn, C.~Adhiambo, K.~Wazny, C.~Ezeaka, E.~Molyneux, M.~Temmerman, P.~Okong, A.~Malata \emph{et~al.}, ``Research priorities in maternal and neonatal health in africa: results using the child health and nutrition research initiative method involving over 900 experts across the continent,'' \emph{AAS open research}, vol.~4, 2021.

\bibitem{tunccalp2017recommendations}
O.~Tun{\c{c}}alp, J.~P. Pena-Rosas, T.~Lawrie, M.~Bucagu, O.~T. Oladapo, A.~Portela, and A.~M. G{\"u}lmezoglu, ``Who recommendations on antenatal care for a positive pregnancy experience-going beyond survival,'' \emph{Bjog}, vol. 124, no.~6, pp. 860--862, 2017.

\bibitem{unicefmaternaldeaths}
\BIBentryALTinterwordspacing
UNICEF, ``Antenatal care is essential for protecting the health of women and their unborn children,'' 2024, accessed: 26 April 2024. [Online]. Available: \url{https://data.unicef.org/topic/maternal-health/antenatal-care/}
\BIBentrySTDinterwordspacing

\bibitem{dusingizimana2023predictors}
T.~Dusingizimana, T.~Ramilan, J.~L. Weber, P.~O. Iversen, M.~Mugabowindekwe, J.~Ahishakiye, and L.~Brough, ``Predictors for achieving adequate antenatal care visits during pregnancy: a cross-sectional study in rural northwest rwanda,'' \emph{BMC Pregnancy and Childbirth}, vol.~23, no.~1, p.~69, 2023.

\bibitem{ahinkorah2021examining}
B.~O. Ahinkorah, E.~K. Ameyaw, A.-A. Seidu, E.~K. Odusina, M.~Keetile, and S.~Yaya, ``Examining barriers to healthcare access and utilization of antenatal care services: evidence from demographic health surveys in sub-saharan africa,'' \emph{BMC health services research}, vol.~21, pp. 1--16, 2021.

\bibitem{mgata2019factors}
S.~Mgata and S.~O. Maluka, ``Factors for late initiation of antenatal care in dar es salaam, tanzania: A qualitative study,'' \emph{BMC Pregnancy and Childbirth}, vol.~19, pp. 1--9, 2019.

\bibitem{maluka2020pregnant}
S.~O. Maluka, C.~Joseph, S.~Fitzgerald, R.~Salim, and P.~Kamuzora, ``Why do pregnant women in iringa region in tanzania start antenatal care late? a qualitative analysis,'' \emph{BMC pregnancy and childbirth}, vol.~20, pp. 1--7, 2020.

\bibitem{smith2019demand}
A.~Smith, R.~Burger, and V.~Black, ``Demand-side causes and covariates of late antenatal care access in cape town, south africa,'' \emph{Maternal and Child Health Journal}, vol.~23, pp. 512--521, 2019.

\bibitem{bobo2021poor}
F.~T. Bobo, A.~Asante, M.~Woldie, and A.~Hayen, ``Poor coverage and quality for poor women: Inequalities in quality antenatal care in nine east african countries,'' \emph{Health Policy and Planning}, vol.~36, no.~5, pp. 662--672, 2021.

\bibitem{sserwanja2022factors}
Q.~Sserwanja, L.~Nuwabaine, G.~Gatasi, J.~N. Wandabwa, and M.~W. Musaba, ``Factors associated with utilization of quality antenatal care: a secondary data analysis of rwandan demographic health survey 2020,'' \emph{BMC health services research}, vol.~22, no.~1, p. 812, 2022.

\bibitem{barron2018mobile}
P.~Barron, J.~Peter, A.~E. LeFevre, J.~Sebidi, M.~Bekker, R.~Allen, A.~N. Parsons, P.~Benjamin, and Y.~Pillay, ``Mobile health messaging service and helpdesk for south african mothers (momconnect): history, successes and challenges,'' \emph{BMJ global health}, vol.~3, no. Suppl 2, p. e000559, 2018.

\bibitem{temgoua2018innovative}
M.~N. Temgoua, J.~N. Tochie, C.~Danwang, V.~M. Aletum, and R.~Tankeu, ``An innovative technology to curb maternal and child mortality in sub-saharan africa: the giftedmomtm approach,'' \emph{Clinical Research in Obstetrics and Gynecology}, vol.~1, no.~1, pp. 1--3, 2018.

\bibitem{obasola2017women}
O.~I. Obasola and I.~M. Mabawonku, ``Women's use of information and communication technology in accessing maternal and child health information in nigeria.'' \emph{African Journal of Library, Archives \& Information Science}, vol.~27, no.~1, 2017.

\bibitem{kaaya2021maternal}
E.~S. Kaaya, J.~Ko, and E.~Luhanga, ``Maternal knowledge-seeking behavior among pregnant women in tanzania,'' \emph{Women's Health}, vol.~17, p. 17455065211038442, 2021.

\bibitem{khamis2023maternal}
S.~Khamis and D.~J. Agboada, ``Maternal health information disparities amid covid-19: Comparing urban and rural expectant mothers in ghana,'' \emph{Media and Communication}, vol.~11, no.~1, pp. 173--183, 2023.

\bibitem{chu2022impact}
J.~N. Chu, U.~Sarkar, N.~A. Rivadeneira, R.~A. Hiatt, and E.~C. Khoong, ``Impact of language preference and health literacy on health information-seeking experiences among a low-income, multilingual cohort,'' \emph{Patient Education and Counseling}, vol. 105, no.~5, pp. 1268--1275, 2022.

\bibitem{massey2017nativity}
P.~M. Massey, B.~A. Langellier, T.~Sentell, and J.~Manganello, ``Nativity and language preference as drivers of health information seeking: examining differences and trends from a us population-based survey,'' \emph{Ethnicity \& health}, vol.~22, no.~6, pp. 596--609, 2017.

\bibitem{khoong2019health}
E.~C. Khoong, G.~M. Le, M.~Hoskote, N.~A. Rivadeneira, R.~A. Hiatt, and U.~Sarkar, ``Health information--seeking behaviors and preferences of a diverse, multilingual urban cohort,'' \emph{Medical care}, vol.~57, pp. S176--S183, 2019.

\bibitem{languagesrwanda2}
\BIBentryALTinterwordspacing
J.~Eyssette, ``Kinyafranglais: how rwanda became a melting pot of official languages,'' 2022, accessed: 13 May 2024. [Online]. Available: \url{https://theconversation.com/kinyafranglais-how-rwanda-became-a-melting-pot-of-official-languages-185441}
\BIBentrySTDinterwordspacing

\bibitem{artieta2018evaluation}
I.~Artieta-Pinedo, C.~Paz-Pascual, G.~Grandes, G.~Villanueva, E.~Q. Group \emph{et~al.}, ``An evaluation of spanish and english on-line information sources regarding pregnancy, birth and the postnatal period,'' \emph{Midwifery}, vol.~58, pp. 19--26, 2018.

\bibitem{virani2019parenting}
A.~Virani, L.~Duffett-Leger, and N.~Letourneau, ``Parenting apps review: in search of good quality apps,'' \emph{Mhealth}, vol.~5, 2019.

\bibitem{storr2017online}
T.~Storr, J.~Maher, and E.~Swanepoel, ``Online nutrition information for pregnant women: a content analysis,'' \emph{Maternal \& child nutrition}, vol.~13, no.~2, p. e12315, 2017.

\bibitem{cannon2020review}
S.~Cannon, M.~Lastella, L.~Vincze, C.~Vandelanotte, and M.~Hayman, ``A review of pregnancy information on nutrition, physical activity and sleep websites,'' \emph{Women and Birth}, vol.~33, no.~1, pp. 35--40, 2020.

\bibitem{al2016online}
B.~H. Al~Wattar, C.~Pidgeon, H.~Learner, J.~Zamora, and S.~Thangaratinam, ``Online health information on obesity in pregnancy: a systematic review,'' \emph{European Journal of Obstetrics \& Gynecology and Reproductive Biology}, vol. 206, pp. 147--152, 2016.

\bibitem{hayman2021quality}
M.~Hayman, K.-L. Alfrey, S.~Cannon, S.~Alley, A.~L. Rebar, S.~Williams, C.~E. Short, A.~Altazan, N.~Comardelle, S.~Currie \emph{et~al.}, ``Quality, features, and presence of behavior change techniques in mobile apps designed to improve physical activity in pregnant women: systematic search and content analysis,'' \emph{JMIR mHealth and uHealth}, vol.~9, no.~4, p. e23649, 2021.

\bibitem{swartzendruber2018contraceptive}
A.~Swartzendruber, R.~J. Steiner, and A.~Newton-Levinson, ``Contraceptive information on pregnancy resource center websites: a statewide content analysis,'' \emph{Contraception}, vol.~98, no.~2, pp. 158--162, 2018.

\bibitem{hamwela2018evaluation}
V.~Hamwela, W.~Ahmed, and P.~Bath, ``Evaluation of websites that contain information relating to malaria in pregnancy,'' \emph{Public health}, vol. 157, pp. 50--52, 2018.

\bibitem{vogels2022sources}
M.~Vogels-Broeke, D.~Daemers, L.~Bud{\'e}, R.~de~Vries, and M.~Nieuwenhuijze, ``Sources of information used by women during pregnancy and the perceived quality,'' \emph{BMC Pregnancy and Childbirth}, vol.~22, no.~1, p. 109, 2022.

\bibitem{ojaperv2023pregnancy}
K.~Ojaperv and S.~Virkus, ``Pregnancy-related health information behaviour of estonian women,'' \emph{Global Knowledge, Memory and Communication}, vol.~72, no.~3, pp. 284--314, 2023.

\bibitem{wexler2020pregnancy}
A.~Wexler, A.~Davoudi, D.~Weissenbacher, R.~Choi, K.~O’Connor, H.~Cummings, and G.~Gonzalez-Hernandez, ``Pregnancy and health in the age of the internet: A content analysis of online “birth club” forums,'' \emph{PloS one}, vol.~15, no.~4, p. e0230947, 2020.

\bibitem{bostwick2019could}
E.~N. Bostwick, D.~Liao, and S.~K. Lee, ``Could i be pregnant? a study of online adolescent pregnancy forums for social support,'' \emph{First Monday}, 2019.

\bibitem{wigginton2017safe}
B.~Wigginton, C.~Gartner, and I.~J. Rowlands, ``Is it safe to vape? analyzing online forums discussing e-cigarette use during pregnancy,'' \emph{Women's Health Issues}, vol.~27, no.~1, pp. 93--99, 2017.

\bibitem{drGoogle}
\BIBentryALTinterwordspacing
M.~Murphy, ``Dr google will see you now: Search giant wants to cash in on your medical queries,'' 2019, accessed: 26 April 2024. [Online]. Available: \url{https://www.telegraph.co.uk/technology/2019/03/10/google-sifting-one-billion-health-questions-day/}
\BIBentrySTDinterwordspacing

\bibitem{asch2019google}
J.~M. Asch, D.~A. Asch, E.~V. Klinger, J.~Marks, N.~Sadek, and R.~M. Merchant, ``Google search histories of patients presenting to an emergency department: an observational study,'' \emph{BMJ open}, vol.~9, no.~2, p. e024791, 2019.

\bibitem{boyd2017applications}
J.~Boyd-Graber, Y.~Hu, D.~Mimno \emph{et~al.}, ``Applications of topic models,'' \emph{Foundations and Trends{\textregistered} in Information Retrieval}, vol.~11, no. 2-3, pp. 143--296, 2017.

\bibitem{zou2018analyzing}
C.~Zou, ``Analyzing research trends on drug safety using topic modeling,'' \emph{Expert opinion on drug safety}, vol.~17, no.~6, pp. 629--636, 2018.

\bibitem{smoll2021barriers}
N.~R. Smoll, J.~Walker, and G.~Khandaker, ``The barriers and enablers to downloading the covidsafe app--a topic modelling analysis,'' \emph{Australian and New Zealand Journal of Public Health}, vol.~45, no.~4, pp. 344--347, 2021.

\bibitem{qin2022exploring}
Z.~Qin and E.~Ronchieri, ``Exploring pandemics events on twitter by using sentiment analysis and topic modelling,'' \emph{Applied Sciences}, vol.~12, no.~23, p. 11924, 2022.

\bibitem{lebena2022preliminary}
N.~Lebe{\~n}a, A.~Blanco, A.~P{\'e}rez, and A.~Casillas, ``Preliminary exploration of topic modelling representations for electronic health records coding according to the international classification of diseases in spanish,'' \emph{Expert Systems with Applications}, vol. 204, p. 117303, 2022.

\bibitem{ni2022topic}
C.~Ni~Ki, A.~Hosseinian-Far, A.~Daneshkhah, and N.~Salari, ``Topic modelling in precision medicine with its applications in personalized diabetes management,'' \emph{Expert Systems}, vol.~39, no.~4, p. e12774, 2022.

\bibitem{lagan2011impact}
B.~M. Lagan, M.~Sinclair, and W.~G. Kernohan, ``What is the impact of the internet on decision-making in pregnancy? a global study,'' \emph{Birth}, vol.~38, no.~4, pp. 336--345, 2011.

\bibitem{lee2022topics}
J.~Y. Lee and E.~Lee, ``What topics are women interested in during pregnancy: exploring the role of social media as informational and emotional support,'' \emph{BMC Pregnancy and Childbirth}, vol.~22, no.~1, p. 517, 2022.

\bibitem{googlemarket}
\BIBentryALTinterwordspacing
statcounter, ``Search engine market share africa,'' accessd: 31 May 2023. [Online]. Available: \url{https://gs.statcounter.com/search-engine-market-share/all/africa}
\BIBentrySTDinterwordspacing

\bibitem{zubiena2022development}
L.~Zubiena, O.~Lewin, G.~Ogunfiditimi, R.~Coleman, J.~Phezulu, T.~Blackburn, and L.~Joseph, ``Development and testing of the health information website evaluation tool (hiwet)--an inter-rater reliability analysis study,'' \emph{Physiotherapy}, vol. 114, pp. e83--e84, 2022.

\bibitem{zubiena2023development}
L.~Zubiena, O.~Lewin, R.~Coleman, J.~Phezulu, G.~Ogunfiditimi, T.~Blackburn, and L.~Joseph, ``Development and testing of the health information website evaluation tool on neck pain websites--an analysis of reliability, validity, and utility,'' \emph{Patient Education and Counseling}, vol. 113, p. 107762, 2023.

\bibitem{blei2003latent}
D.~M. Blei, A.~Y. Ng, and M.~I. Jordan, ``Latent dirichlet allocation,'' \emph{Journal of machine Learning research}, vol.~3, no. Jan, pp. 993--1022, 2003.

\bibitem{lee1999learning}
D.~D. Lee and H.~S. Seung, ``Learning the parts of objects by non-negative matrix factorization,'' \emph{nature}, vol. 401, no. 6755, pp. 788--791, 1999.

\bibitem{o2015analysis}
D.~O’callaghan, D.~Greene, J.~Carthy, and P.~Cunningham, ``An analysis of the coherence of descriptors in topic modeling,'' \emph{Expert Systems with Applications}, vol.~42, no.~13, pp. 5645--5657, 2015.

\bibitem{latif2021analyzing}
S.~Latif, F.~Shafait, R.~Latif \emph{et~al.}, ``Analyzing lda and nmf topic models for urdu tweets via automatic labeling,'' \emph{IEEE Access}, vol.~9, pp. 127\,531--127\,547, 2021.

\bibitem{kherwa2019topic}
P.~Kherwa and P.~Bansal, ``Topic modeling: a comprehensive review,'' \emph{EAI Endorsed transactions on scalable information systems}, vol.~7, no.~24, 2019.

\bibitem{gugsa2016newspaper}
F.~Gugsa, E.~Karmarkar, A.~Cheyne, and G.~Yamey, ``Newspaper coverage of maternal health in bangladesh, rwanda and south africa: a quantitative and qualitative content analysis,'' \emph{BMJ open}, vol.~6, no.~1, p. e008837, 2016.

\bibitem{conrad2024health}
M.~Conrad, ``Health information-seeking internet behaviours among pregnant women: a narrative literature review,'' \emph{Journal of Reproductive and Infant Psychology}, vol.~42, no.~2, pp. 194--208, 2024.

\bibitem{rurangirwa2018quality}
A.~A. Rurangirwa, I.~Mogren, J.~Ntaganira, K.~Govender, and G.~Krantz, ``Quality of antenatal care services in rwanda: assessing practices of health care providers,'' \emph{BMC health services research}, vol.~18, pp. 1--10, 2018.

\bibitem{mniszak2020nothing}
C.~Mniszak, H.~L. O'Brien, D.~Greyson, C.~Chabot, and J.~Shoveller, ``“nothing's available”: Young fathers’ experiences with unmet information needs and barriers to resolving them,'' \emph{Information Processing \& Management}, vol.~57, no.~2, p. 102081, 2020.

\bibitem{onyeze2020experiences}
C.~Onyeze-Joe and I.~Godin, ``Experiences, views and needs of first-time fathers in pregnancy-related care: a qualitative study in south-east nigeria,'' \emph{BMC pregnancy and childbirth}, vol.~20, pp. 1--11, 2020.

\bibitem{nuwabaine2023sexual}
L.~Nuwabaine, J.~Kawuki, E.~Amwiine, J.~B. Asiimwe, Q.~Sserwanja, G.~Gatasi, E.~Donkor, and H.~Atwijukiire, ``Sexual violence and associated factors among women of reproductive age in rwanda: a 2020 nationwide cross-sectional survey,'' \emph{Archives of public health}, vol.~81, no.~1, p. 112, 2023.

\bibitem{bahati2021intimate}
C.~Bahati, J.~Izabayo, J.~Niyonsenga, V.~Sezibera, and L.~Mutesa, ``Intimate partner violence as a predictor of antenatal care services utilization in rwanda,'' \emph{BMC pregnancy and childbirth}, vol.~21, pp. 1--11, 2021.

\bibitem{nkurunziza2024integrating}
A.~Nkurunziza, V.~L. Smye, C.~N. Wathen, K.~T. Jackson, D.~F. Cechetto, P.~Tryphonopoulos, and D.~Gishoma, ``Integrating trauma-and violence-informed care for adolescent mothers in rwanda: a qualitative study with community health workers,'' \emph{BMC health services research}, vol.~24, no.~1, p. 868, 2024.

\bibitem{world2022guide}
W.~H. Organization, \emph{Guide for integration of perinatal mental health in maternal and child health services}.\hskip 1em plus 0.5em minus 0.4em\relax World Health Organization, 2022.

\bibitem{charnock1999discern}
D.~Charnock, S.~Shepperd, G.~Needham, and R.~Gann, ``Discern: an instrument for judging the quality of written consumer health information on treatment choices.'' \emph{Journal of Epidemiology \& Community Health}, vol.~53, no.~2, pp. 105--111, 1999.

\bibitem{namoun2021review}
A.~Namoun, A.~Alrehaili, and A.~Tufail, ``A review of automated website usability evaluation tools: Research issues and challenges,'' in \emph{International Conference on Human-Computer Interaction}.\hskip 1em plus 0.5em minus 0.4em\relax Springer, 2021, pp. 292--311.

\bibitem{alexander2021influence}
R.~Alexander, N.~Thompson, T.~McGill, and D.~Murray, ``The influence of user culture on website usability,'' \emph{International journal of human-computer studies}, vol. 154, p. 102688, 2021.

\bibitem{stige2024artificial}
{\AA}.~Stige, E.~D. Zamani, P.~Mikalef, and Y.~Zhu, ``Artificial intelligence (ai) for user experience (ux) design: a systematic literature review and future research agenda,'' \emph{Information Technology \& People}, vol.~37, no.~6, pp. 2324--2352, 2024.

\end{thebibliography}

\end{document}